\documentclass[a4paper,aps,prd,10pt,showpacs,showkeys]{revtex4}
\usepackage{graphicx,amsmath,amsfonts,amssymb,revsymb, 
epsfig,bm,xspace,float}
\usepackage{color}
\usepackage{bbold}
\usepackage{subfigure}

\def\beq{\begin{eqnarray}}
\def\eeq{\end{eqnarray}}

\renewcommand{\vec}[1]{{\bf #1}}

\def\al{\alpha}
\def\be{\beta}

\def\vp{\varepsilon}

\def\ka{\kappa}
\def\la{\lambda}
\def\na{\nabla}
\def\pa{\partial}

\def\si{\sigma}
\def\om{\omega}

\def\Ga{\Gamma}

\def\La{\Lambda}

\def\n{\label}                          
\def\r{\ref}                             

\newcommand{\lp}{\ensuremath{\left(}}
\newcommand{\rp}{\ensuremath{\right)}}

\newcommand{\MP}{M_\mathrm{_P}}

\begin{document}

\title{Recent progress in fighting ghosts in quantum gravity}

\author{Filipe de O. Salles}
\email{salles@cbpf.br}
\affiliation{Centro Brasileiro de Pesquisas F\'isicas,
22290-180 URCA, Rio de Janeiro (RJ), Brazil}

\author{Ilya L. Shapiro\footnote{Based on the invited talk given
by I.Sh. at the quantum gravity meeting in SUSTech, Shenzhen, China. }}
\email{shapiro@fisica.ufjf.br}
\affiliation{Departamento de F\'{\i}sica, ICE, Universidade Federal de Juiz de Fora, 36036-900, MG, Brazil
\\
Tomsk State Pedagogical University, Tomsk, 634041, Russia
\\
Tomsk State University, Tomsk, 634050, Russia}

\keywords{Higher derivatives, massive ghosts, stability,
cosmological solutions, gravitational waves.}

\pacs{
04.62.+v,  
98.80.-k,  
04.30.-w   
}

\begin{abstract}
We review some of the recent results which can be useful for
better understanding of the problem of stability of vacuum and in
general classical solutions in higher derivative quantum gravity.
The fourth derivative terms in the purely gravitational vacuum
sector are requested by renormalizability already in both
semiclassical and complete quantum gravity theories. However,
because of these terms the spectrum of the theory has unphysical
ghost states which jeopardize the stability of classical solutions.
At the quantum level ghosts violate unitarity, and and thus
ghosts look incompatible with the consistency of the theory. The
``dominating'' or ``standard'' approach is to treat higher derivative
terms as small perturbations at low energies. Such an effective
theory is supposed to glue with an unknown fundamental theory
in the high energy limit. We argue that the perspectives for such
a scenario are not clear, to say the least. On the other hand,
recently there was certain progress in understanding physical
conditions which can make ghosts not offensive. We survey these
results and discuss the properties of the unknown fundamental
theory which can provide these conditions satisfied.
\end{abstract}

\maketitle

\section{Introduction}

Numerous tests and verifications performed during the last century
have shown that General Relativity
(GR) is a complete theory of classical gravitational phenomena. GR
proved valid and useful in the wide range of energies and distances.
At the same time, the presence of singular regions in physically
relevant solutions of GR indicates the need for extending the theory.
One can assume that GR is not valid at all scales, especially at very
short distances and/or when the curvature becomes very large.
In this situation one can expect that the gravitational phenomena
should be described by a more extensive and complicated theory.
Indeed, one should expect that this unknown theory coincides with
GR at large distances and/or in the weak field limit.

The dimensional arguments indicate that the origin of deviations from
GR is most likely related to quantum effects. E.g., the existence of
fundamental Planck units ($M_{\rm P} \sim 10^{19}\,GeV$)
hints the possibility of some sort of a new fundamental physics at
the very high energy scale, where relativistic, quantum and
gravitational effects become relevant at the same time. How can
we interpret such a result of the dimensional analysis?

One can introduce a simple general classification of all
possible approaches to Quantum Gravity (QG), which is
based on the object of quantization.  There are three
distinct groups of approaches, namely
\vskip 1mm

{\it i)} \
Quantize both gravity and matter fields. This is, definitely,
the most fundamental possible approach.
\vskip 1mm

{\it ii)} \
Quantize only matter fields on classical curved background
(semiclassical approach). This is, in some sense, the most
important approach, since we know for sure that matter
fields should be quantized on a curved background. The main
question is what is the effect (back-reaction) of such a quantum
theory on the gravitational equations of motion.
\vskip 1mm

{\it iii)}
\ Quantize ``something else''. E.g., in case of (super)string theory
both matter and gravity are induced, and the fundamental
object of quantization is the two-dimensional (2D) string, which
lives in the external $D$-dimensional background and defines its
geometry and dynamics.
\vskip 2mm

Which of these approaches is ``better''? The final verdict can be
achieved only in experiments, and purely theoretical arguments
can only help us to select what we regard more consistent, simple
and natural. On the other hand, all these approaches have something
in common, namely there are higher derivative terms in the
gravitational action in all cases. In the next section we briefly
consider this issue in the framework of semiclassical approach.
After that in Sec.~\ref{s3} we discuss that very similar situation
takes place in the theory of quantum gravity and also is quite similar
in string theory. Starting from Sec. \ref{s4} we review the original
results of Refs.~\cite{HD-Stab,Salles:2017xsr} concerning recent
advances in exploring the unitarity of quantum theory in the presence
of complex conjugate pairs of  higher derivative ghosts and in the
study of stability on the cosmological backgrounds.

\section{Semiclassical approach and higher derivatives}
\label{s2}

Without quantization of gravity, at the quantum level the classical
action of vacuum is replaced by the effective action, that includes
contributions of quantum matter fields $\Phi$
\cite{birdav,book} (see also \cite{PoImpo}
for a more recent review),
\beq
e^{i \Ga(g_{\mu\nu})}
\,=\, e^{i S_{vac}(g_{\mu\nu})}\,
\int d\Phi\,e^{i S_m(\Phi,\,g_{\mu\nu})}.
\eeq
The form of the classical action of vacuum is defined by the
consistency conditions, this means that the theory should be
renormalizable. The simplest minimal vacuum action of
renormalizable quantum field theory (QFT) in curved space is
\beq
S_{\rm vac}=S_{EH}+S_{HD},
\eeq
where
\beq
S_{EH}=-\frac{1}{16\pi G}
\int d^4x\sqrt{-g}\,\left\{R + 2\La \right\}
\eeq
is the Einstein-Hilbert action with the cosmological constant and
\beq
S_{HD}=\int d^4x\sqrt{-g}\left\{
a_1C^2 + a_2E_4 + a_3{\Box}R + a_4R^2\right\}
\eeq
includes fourth derivatives, e.g.,  is the square of the Weyl tensor
and
\beq
E_4 =R_{\mu\nu\al\be}^2 - 4R_{\al\be}^2 + R^2
\n{GB4}
\eeq
is the integrand of the Gauss-Bonnet topological term.

Without higher derivative (HD) terms in the vacuum sector the
semiclassical theory is not consistent due to the non-renormalizability.
Even if these terms are not included into the classical action, they will
emerge due to the renormalization group running in quantum theory.
This can be explicitly seen using the conformal anomaly, as discussed
in \cite{PoImpo,GWprT}. Formally, regarding semiclassical theory
as fundamental  (not effective) QFT, the higher derivative terms are
not quantum corrections, for they should be introduced already at the
classical level.

\section{Two sides of higher derivatives in quantum gravity}
\label{s3}

Consider now the situation in QG. The  renormalizability of QG
models strongly depend on the choice of the initial classical action.
As the first example, let us consider quantum GR.
\beq
S_{EH} \,=\, -\,\frac{1}{16\pi G}\int d^4x \sqrt{-g} \,(R + 2\La)\,.
\eeq
Using the standard power counting arguments (see some details
below) one can easily obtain the relation
\beq
D+d\,=\,2+2p,
\n{indexGR}
\eeq
where $D$ is the superficial degree of divergence of a diagram
with $p$ loops and $d$ is the number of derivatives acting on the
external lines of the diagram. One can easily see from (\r{indexGR})
and covariance of the counterterms that at the 1-loop level
there are logarithmically divergent term which are quadratic in
curvatures~\cite{hove,dene}, namely
\beq
{\cal O}(R^2_{...}) = R_{\mu\nu\al\be}^2,\quad
R_{\mu\nu}^2,\quad
R^2,\quad
\Box R.
\eeq
At the 2-loop level we have \cite{gorsag},
\beq
{\cal O}(R^3_{...}) = R_{\mu\nu}\Box R^{\mu\nu}\,,\,...\,
R^3\,,\,\,\,
R_{\mu\nu}R^{\mu}_\al R^{\al\nu}\,,\,\,\,
R_{\mu\nu\al\be}R^{\mu\nu}\,_{\rho\si} R^{\mu\nu\rho\si}\,.
\eeq
Since the last of these structures does not vanish on-shell, the theory
is not renormalizable in the usual sense. Of course, one can rely of
the effective approach and make sound calculations (see, e.g.,
\cite{Don94} and the review \cite{Burgess}), but the approximation
behind this approach breaks down at the Planck scale, where QG is
supposed to be especially relevant.

Within the standard perturbative approach non-renormalizability means
the theory has no predictive power. Every time we introduce a new type
of a counterterm, it is necessary to fix renormalization condition and
this means a measurement. So, before making a single predictions, it
is necessary to have an infinite amount of experimental data.

What are the possible solutions of this problem? One of the options
is to trade the standard perturbative approach in QFT to something
different. Another way out is to modify or generalize the theory, i.e.,
start from another theory to construct QG. The first option is widely
explores in the asymptotic safety scenarios, in the effective
approaches to QG  (which was mentioned above), induced gravity
approach (including string theory) and so on. Regardless of many
options, their consistency and relation to the general targets of the
QG program are not completely clear, in all cases. In what follows
we shall concentrate on the second possibility and consider
modified action of gravity as a starting point to construct QG.

The most natural choice is start from the four derivative gravity
model,  because we need fourth derivatives anyway to deal with
the quantum matter field. Then the starting action should be
\beq
S_{\rm gravity}\,=\,S_{EH}\,+\,S_{HD},
\n{gravity1}
\eeq
where $S_{EU}$ is the Einstein-Hilbert action (as mentioned before)
and $S_{HD}$ includes square of the Weyl tensor and $R$,
\beq
S_{HD} = - \int d^4x\sqrt{-g}\left\{
\frac{1}{2\la}\,C^2
+ \frac{\om}{3\la}\,R^2
\,+\,\mbox{ surface terms}\right\}.
\eeq
The propagators of metric and ghosts behave like ${\cal O}(k^{-4})$	
(in the notations of \cite{Stelle} and \cite{book} this means $r_l=4$)
and we have $K_4$, $K_2$, $K_0$ vertices with four, two and
zero powers of momenta. The superficial degree of divergence $D$
of the diagram with an arbitrary number of loops satisfies the relation
\beq
D+d=4-2K_2-4K_0,
\eeq
where $d$ is the number of derivatives of external metric lines.
So, this theory is definitely renormalizable and the dimensions of
possible counterterms are 4,\,\,2,\,\,0, depending on number of
vertices with lower derivatives \cite{Stelle}.

However, one has to pay a very high price for renormalizability,
since this theory has massive ghosts. This can be seen from the
spin-two sector of the propagator \cite{Stelle},
\beq
G_{\rm spin-2}(k)\,\sim\,\frac{1}{m^2}\,\,\Big(
\frac{1}{k^2} - \frac{1}{k^2+m_2^2} \Big),
\qquad
\mbox{where}
\qquad
m_2 \propto M_P\,.
\eeq
The tree-level spectrum includes massless graviton and massive
spin-$2$ ``ghost'' with negative kinetic energy and a huge mass.
The presence of a particle with negative energy means possible
instability of the vacuum state of the theory. For instance, the
Minkowski space is not protected from the spontaneous creation
of massive ghost and (needed for energy conservation)
compensating gravitons from the vacuum.

Indeed, there are different sides of the High Derivative
Quantum Gravity (HDQG) problems with massive ghosts. For
instance,
\vskip 1mm

{\it i)} \
In classical systems higher derivatives may generate exploding
instabilities at the non-linear level \cite{Ostrog}
(see, e.g., recent review in \cite{Woodard-r}).
\vskip 1mm

{\it ii)} \
Interaction between massive unphysical ghost and gravitons leads to
massive emission
of gravitons and unbounded acceleration of ghost. As a result one
should observe violation of energy conservation in the massless
sector \cite{Veltman-63}, that means an explosion of gravitons.
Also, ghosts produce violation of unitarity of the S-matrix,
which also means similar instability at the quantum level.
\vskip 2mm

Due to the great importance of the problem of higher derivatives
and ghosts, there was many proposals to solve it, e.g.,
\cite{Tomb-77,salstr} and \cite{Hawking}. Let us consider another
proposal, related to further generalization of the action of the QG
theory. One can include more than four derivatives \cite{highderi},
\beq
&&
S = S_{EH}
\,+\,
\sum\limits_{n=0}^{N}
\int d^4x\sqrt{-g}\Big\{
\om_n^C C_{\mu\nu\al\be} \Box^n C_{\mu\nu\al\be}
+ \om_n^R R \Box^n R
\Big\} + {\cal O}\big(R^3_{\dots}\big).
\n{SuperAct}
\eeq
A simple analysis shows that in this theory massive ghost-like states
are still present. For the real poles case we can write
\beq
G_2(k) = \frac{A_{0}}{k^2} +
 \frac{A_{1}}{k^2 + m_1^2} + \frac{A_{2}}{k^2 + m_2^2}
+ \cdots + \frac{A_{N+1}}{k^2 + m_{N+1}^2},
\eeq
and it has been shown  \cite{highderi} that for any sequence of poles
with   $\,0 < m_1^2 < m_2^2 < m_3^2 < \cdots < m_{N+1}^2$,
the signs of the corresponding terms alternate, $A_j\cdot A_{j+1} < 0$.
This means that one can not make all but one particle in the spectrum
to be healthy and provide an infinite mass of the ghost. In this sense
the theory (\r{SuperAct}) has the same level of problems with ghosts
that the simpler fourth-derivative model.

However, the renormalization properties of these two theories are
quite different. It is easy to see that the theory  (\r{SuperAct}) is
superrenormalizable if both higher order terms are present,
\ $\om_N^C \cdot \om_N^R \neq 0$. In
order to check this fact, consider the power counting in this case.
For the sake of simplicity we can consider only the vertices with
a maximal number \ $K_\nu$ \ of  maximal derivatives, $r_l=2N+4$,
which obviously provide the maximal power of divergences.

The propagators of gravitational modes and ghosts in this model
are ${\cal O}(k^{-r_l})$, where and, combining the
general expression for power counting for the diagram with \ $n$ \
vertices and \ $p$ \ loops,
\beq
D + d \,=\,\sum\limits_{l_{int}}(4-r_l)
\,-\,4n \,+\,4\,+\,\sum\limits_{\nu}K_\nu
\eeq
with the topological relation for the number of internal lines,
\beq
\qquad
l_{int} = p + n - 1,
\eeq
one can easily arrive at the estimate of $d$ for the logarithmic
divergences with $D=0$,
\beq
d\,=\,4\,+\,N(1-p)\,.
\eeq
For $N=0$ we meet the standard HDQG result, $d=4$. Due to the
covariance, this means that the counterterms repeat the form of the
four-derivative action $\,S_{\rm gravity}\,$ in Eq.~(\r{gravity1}).
It is remarkable that the terms with six and higher derivatives do
not get renormalized, but the coefficients of these terms define
the divergences. Starting from $N=1$ we have superrenormalizable
theory, where the divergences show up only in $p=1,2,3$ loops.
For $N\geq 3$ we have such a superrenormalizable theory, where
divergences exist only for $p=1$, that is at the one-loop level. Let
us stress that the one-loop divergences are present for all $N$ and
that the logarithmic divergences always have zero, two and four
derivatives of the metric, independent on $N$.

The low-energy effects of complex and real ghosts in these models
were recently discussed in \cite{Newton-high,ABS}.
Another interesting possibility is that one can derive {\it exact}
$\be$-functions in this superrenormalizable QG model, by means of
one-loop level calculations \cite{highderi,SRQG-beta}. These calculations,
anyway, may be very difficult and for a while the results were achieved only
for the beta functions of cosmological and Newton constants. They have
the form
\beq
\beta_{\La}
&=& \mu \frac{d \rho_\La}{d\mu}
= \frac{1}{(4\pi)^2}
\Big(
\frac{5 \omega _{N-2,C}}{\omega _{N,C}}
+\frac{\omega _{N-2,R}}{\omega _{N,R}}
-\frac{5 \omega _{N-1,C}^2}{2 \omega _{N,C}^2}
-\frac{\omega _{N-1,R}^2}{2 \omega_{N,R}^2} \Big),
\qquad
\,\rho_\La = \frac{\La}{8\pi G};
\\
\beta_{G}
&=& \mu \frac{d}{d\mu}\,\Big( - \frac{1}{16 \pi G}\Big)
= -  \frac{1}{6(4\pi)^2}\,
\Big(\frac{5\om_{N-1,C}}{\om_{N,C}}
+ \frac{\om_{N-1,R}}{\om_{N,R}}\Big).
\eeq
Here we used the standard notation for the density of the cosmological
constant $\,\rho_\La$.

Different from four-derivative quantum gravity these $\be$-functions
do not depend on the choice of a gauge-fixing condition
\cite{highderi,SRQG-beta}. This important feature follows from
the fact that the classical equations of motion and the the divergences
in this theory have different number of metric derivatives. And, once
again, for $N \geq 3$ these universal beta-functions are exact.

All in all, one can see that from the theoretical side there the positive
and negative aspects of introducing the higher derivative terms in
quantum gravity.
The consistent theory which is supposed to work at arbitrary energy
scale can not be constructed without at least fourth derivatives.
If the higher derivative terms are included, then the tree-level
spectrum will include massless graviton and massive spin-2 ``ghost''
with negative kinetic energy and huge mass. If we do not include
the higher derivative terms into classical action, they will emerge
with infinite coefficients and (most relevant) with logarithmically
running parameters. In any case, the nonphysical ghosts come back.

Thus, we can reach the following general conclusion: there is no
way to live with ghosts and, on the other hand, there is no way
to live without ghosts. The situation looks like a strange puzzle.
However, parallel to this strange conclusion there is one
absolutely certain thing. As a matter of fact the world exists, we
live, and so there must be some explanation and resolution of
the mentioned puzzle, of course.

The standard (for some people, at least) logic to solve this issue
is to consider, {\it by definition}, all higher derivative terms to
be small perturbations \cite{Simon-90,parsim}.
In this approach all higher derivative terms, including the terms
in the classical action which are subject of renormalization,
local and nonlocal quantum corrections, running parameter etc,
are regarded as small perturbations over the basic Einstein-Hilbert
term of GR. Certainly, this approach is efficient in fighting ghosts.
However, a bad news is that it is completely {\it ad hoc} approach.
Furthermore, it is based on the approximation which is efficient
only for the energies which are much below the Planck scale.
And this is not what we expect from the ``theory of everything'',
such as QG. As far as we approach the Planck energies, the
higher derivative terms can not be treated as small. Another
disadvantage is that this {\it ad hoc} procedure brings a lot of
ambiguity. For instance, how should we treat the $R^2$ term?
Taking it as perturbation is somehow groundless, since it does
not make ghosts. At the same time, from the dimensional and
conceptual viewpoints there is no apparent difference between
$R^2$ and $R_{\mu\nu}^2$ terms, so why they should be
treated different? And worst than that, treating $R^2$ term
as perturbation, we are forced to ``forbid'' the Starobinsky
model of inflation, which is phenomenologically very
successful. Let us stress that this inflationary model is essentially
based on treating $R$ and $R^2$ terms at the equal level, and not
taking the last one as a perturbation.

Another important issue is what to do with $R^3$, \
$R R_{\mu\nu}R^{\mu\nu}$, and other similar terms.
Why should we treat all such terms as perturbations?
Because they have higher derivatives? Even regardless of
the fact they do not produce ghosts? What is the rule of
splitting the action into the main part and perturbation?

We may think that if the criterium is dimension, then this
approach means that we assume that quantum gravitational
phenomena are relevant only far below the Planck scale.
And, let us repeat, this is something opposite to what we
expect from QG, since the original motivation was to deal
with the Planck energies.

\section{
Ghosts in string theory and in the non-polynomial Quantum Gravity}
\label{s4}

Let us consider two examples of ghost-free HD models of gravity.
Both models can be seen as different representations of string or
superstring theory. In string theory, the object of quantization is a kind
of non-linear sigma-model in two space-time dimensions. In this case
both metric and matter fields are induced, implying unification of all
fundamental forces. The sigma-model approach to string theory (we
consider only bosonic case) is a QFT in $2D$ curved space,
\beq
S_{\rm string} &=& \int {d^2}\sigma\sqrt {g} \left\{ \frac{1}{2 \alpha'}
g^{\mu\nu} G_{ij}(X)
\pa_{\mu}{X^i} \pa_{\nu}{X^j}
\right.
\\
\nonumber
&+&
\left.
\frac{1}{\alpha'}\,
\frac{\vp^{\mu\nu}}{\sqrt {g}}\, {A_{ij}}(X)
\partial_{\mu} {X^i} \partial_{\nu} {X^j} + B(X)R + T(X)
\right\}\,,
\quad
i,j = 1,2,...,D\,.
\n{str}
\eeq
In the Polyakov approach the conditions of anomaly cancellation
emerge order by order in ${\alpha'}$. This expansion corresponds
to the special order of functional integration and to the low-energy
effective action which corresponds to the growing orders of
metric derivatives \cite{BSSh-NPB}. The critical dimensions are
\begin{center}
D=26  for bosonic string
\quad
and
\quad
D=10 for superstrings.
\end{center}
At the first order in ${\alpha'}$ the effective equations give
induced GR \cite{FrTs-85,CFMP}, coming from the condition
of Weyl invariance of string at the quantum level.
In the second order in  ${\alpha'}$ the low-energy effective
action already has the same fourth order in derivatives terms,
which we already met in QG. However, in string theory there
one extra possibility. Namely, one can use special reparametrization
of the metric $G_{\mu\nu}$ to remove ghosts at all orders in
${\alpha'}$. In the simplest torsionless case the effective action
can can be written as
\beq
S_M=\frac{2}{\kappa^2}\int d^Dx \sqrt{G}\; e^{-2\phi}\,
\Big\{ -R + 4\,(\partial
\phi)^2
+ \alpha'\,\big( a_1R_{\lambda\mu\nu\rho}R^{\lambda\mu\nu\rho}
+ a_2R_{\mu\nu}R^{\mu\nu}+a_3R^2\big)\Big\} + \,\,..\,,
\eeq
where the dilaton $\phi$ is related to the $B(X)$ in Eq.~(\r{str}).
Now, in order to remove ghosts one performs reparametrization
of the background metric $G_{\mu\nu}$ as follows:
\beq
G_{\mu\nu} \longrightarrow
G'_{\mu\nu} =
G_{\mu\nu} +
\al '\left(x_1\,R_{\mu\nu} + x_2\,R\, G_{\mu\nu}\right) + ...\,,
\n{killghost}
\eeq
where $x_{1,2,...}$ are specially tuned parameters \cite{zwei}.

It is important to note that the reparametrization (\r{killghost})
doesn't affect string $S$-matrix, because it does not concern
quantum fields \cite{zwei}.
\
At the same time the coefficients $\,x_1,\,x_2,\,x_3, \dots$ can
be chosen in such a way that the effective low-energy theory of
metric becomes free of massive unphysical ghosts. For instance,
the fourth derivative terms combine into the Gauss-Bonnet
term (\r{GB4}), namely
\beq
\int d^Dx \sqrt{G}\,
\Big\{ R_{\lambda\mu\nu\rho}R^{\lambda\mu\nu\rho}
- 4 R_{\mu\nu}R^{\mu\nu} + R^2\Big\},
\n{GB}
\eeq
which is topological for $4D$ but does not contribute to the
propagator in any space-time dimension $D$. The same is true
for the combination with extra factors of $\Box$,
\beq
\int d^Dx \sqrt{G}\,
\Big\{ R_{\lambda\mu\nu\rho}\Box^n R^{\lambda\mu\nu\rho}
- 4 R_{\mu\nu}\Box^n  R^{\mu\nu} + R\Box^n  R\Big\},
\n{GB-box}
\eeq
which may be  achieved in the higher orders in ${\alpha'}$ by
correctly tuning higher order coefficients $x_{3,4,\dots}$.
As a result the theory of string produces induced gravity which
is free of ghosts and has no issues with renormalizability, since
gravity is all induced. All this means that string theory solves the
problem of QG in a satisfactory way, of course if we believe that
gravity should be induced from string.

It is worthwhile, however, to look into further details of the
scheme described above. The first observation is that the
reparametrization (\r{killghost}) is ambiguous and this actually
produce ambiguous physical solutions, e.g., in cosmology
\cite{maroto}. For instance, the terms of the form $f(R)$
can be arbitrarily changed or removed by this transformation,
and this ambiguity and, in general, $f(R)$, do not affect the
presence of ghosts at all. One can
note, for example, that the most successful model of inflation
by Starobinsky \cite{star} requires the $R^2$ term with the
well-defined coefficient. Then we have to tune the parameter
$x_2$ in  (\r{killghost}) such that after the compactification of
extra dimensions one can provide this desirable value of the
coefficient of $R^2$, instead of making it zero.

Even more subtle point is that the effectively working
ghost-killing transformation (\ref{killghost}) must be
{\it absolutely precise.}
An infinitesimal change in the fine-tuning of the parameters
\ $x_{1,2,3,4,\dots}$ \ would immediately create a ghost with
a huge mass.
Moreover, smaller violation of the absolutely precise
fine-tuning leads to a greater mass of the ghost, hence (according
to a ``standard wisdom'') smaller violation of fine-tuning produce
greater gravitational instability.

Furthermore, we know from all our experience in Physics that
at low energies quantum effects are described, e.g., by QFT, and
not by the string theory. Even higher loop corrections in QED
eventually lead to the small violation of the  absolutely precise
ghost-killing transformation (\ref{killghost}) and produce
a huge destructive ghost, as we explained above. Hence, string
theory is ghost-free and unitary theory of QG, but only if it
completely controls all QFT effects, even in the deep IR. It
means that string theory must be a real and complete theory of
everything, in order to be a consistent theory of QG. The reality
of such a control is not obvious, in our opinion.

The second example is an interesting alternative to the original
Zwiebach transformation (\ref{killghost}).  In the non-local
theory \cite{Tseytlin-95}
\beq
S = -\, \frac{1}{2\ka}
\int d^4 x \sqrt{-g}\,\,
\Big\{ R
+ \, G_{\mu\nu} \,\frac{ a(\Box) -1}{\Box}\, R^{\mu\nu}
\Big\}\,,
\quad
a (\Box) = e^{-\Box/m^2}\,.
\n{nonlo}
\eeq
there are no ghosts, regardless of the presence of infinite
derivatives in the action (an interesting discussion of physical
spectrum and Cauchy problem in the theories of this kind
was recently given in \cite{CM}).

In this and similar theories propagator of metric perturbations
has a single massless pole, corresponding to gravitons. With
this choice there are no ghosts. The idea is to use Zwiebach-like
transformation (\ref{killghost}), but arrive at the non-local theory
(\r{nonlo}) which is, non-polynomial in derivatives, instead  of
``killing'' all higher derivatives that one can kill. From the
viewpoint of string theory this means we have one more
ambiguity in the effective low-energy action of gravity.

However, the same action can be used in a distinct way. There was a
proposal to use the same kind of non-local models to construct
superrenormalizable and unitary models of QG
\cite{Tomboulis-97,Modesto2011}. In such a theory the propagator
is defined by the terms bilinear in curvature,
\beq
S = \int\limits_x \Big\{ - \frac{1}{\ka^2} \,R
+ \frac12\,C_{\mu\nu\al\be}  \, \Phi(\Box) \, C^{\mu\nu\al\be}
+ \frac12\,R \, \Psi (\Box)\, R\Big\}\,.
\eeq
The equation for defining the poles is,
\beq
p^2\,\Big[1 \,+\, \ka^2 p^2\Phi(-p^2)\Big]
\,=\, p^2\,e^{\al p^2}\,=\,0.
\eeq
In this particular case there is only a massless pole corresponding
to gravitons. But unfortunately, it is impossible to preserve the
ghost-free structure at the quantum level \cite{CountGhost}.
Typically, after taking the loop corrections into account, in the
dressed propagator there are infinitely many poles on the complex
plane. In this sense the ghost-free structure of the theory can not
be preserved beyond the tree level.

So we can make a conclusion that there is no way to live without
ghosts in QG. In all three fundamental approaches to QG, namely
semiclassical, legitimate QG, and induced gravity/strings, there is
no reasonable way to get rid of massive ghost-like states.

At this stage we can only repeat that there is apparently no way
to live with ghosts, since their presence implies
instability of all classical gravitational solutions and violation of
unitarity. In other words, at both classical and quantum level
ghosts do not enable one to have a consistent theory. Therefore
we have a deep conflict between renormalizability and
unitarity/stability. At the moment there is no solution of this
great puzzle, but in what follows we present some recent
advances in its better understanding.

\section{Complex poles: old expectations in the new setting}
\label{s5}

The importance of higher derivatives in semiclassical and
quantum gravity has been fully recognized in the early 60-ies
\cite{UtDW}, and the bad features of ghosts was completely
clear more than 50 years ago \cite{Stelle}. In the time period
which passed after that there were numerous proposal on
solving the contradiction between renormalizability and
unitarity in QG. In particular, there was a promising idea that
ghosts may become complex after taking the loop contributions
into account. This means that there can be only complex
``massive'' poles in the dressed propagator \cite{Tomb-77,salstr}.
Such poles always come in complex conjugate pair, which opens
interesting possibilities, related to the Lee-Wick quantization
scheme (let us note that another, different approach to deal
with ghosts has been  suggested by Hawking and Hertog
in \cite{Hawking}). Similar approaches to solve the problem
of higher derivative massive ghosts in fourth derivative QG
were discussed in \cite{Antomb} and finally reviewed in
\cite{Johnston}. In the last reference it was shown that the
definitive answer on whether this mechanism works can be
obtained only on the basis of the full non-perturbative dressed
propagator of the gravitational perturbations. One-loop
effects of matter fields and proper gravity, large-$N$
approximation and lattice-based considerations indicated
an optimistic picture, but unfortunately all of these results
are not conclusive, as explained in \cite{Johnston}.
As far as we do not have completely reliable nonperturbative
approach to QG, the chances to get a complete information
about the exact dressed propagator look rather remote
(let us mention an interesting attempt \cite{RobertoP} to use
Functional Renormalization group method for this end).
But, do we always need so much to analyse the structure of
the dressed propagator?

Starting from \cite{salstr} and \cite{Tomb-77}, the
main hope for the ``minimal'' fourth-derivative QG was that the
real ghost pole splits into a couple of complex conjugate poles
under the effect of quantum corrections. And we can not
control the position of these complex poles in the dressed
propagator, since the higher loop corrections can be
complicated, essential and difficult to evaluate.  However,
for the theory of QG with six or more derivatives \cite{highderi}
all this is not necessary at all! In this case one can simply start
from the tree-level theory which has complex conjugate massive
poles from the very beginning, and hence there is no need to
rely on the precise knowledge of a dressed propagator. In this
way one can successfully construct the theory of quantum gravity
which is
both unitary and superrenormalizable \cite{LQG-D4} (see also
generalization for an arbitrary dimension in \cite{Modesto2016}).

Furthermore, one can prove that in this models the unitary holds
also at the quantum level, in particular because in such a
superrenormalizable model one can guarantee that the position
of the poles in the dressed propagator will be qualitatively
the same as in the tree level theory. Further features of this kind
of models, such as reflection positivity, has been discussed
recently in Refs.~\cite{ARS,MOD}, with somehow contradicting
results. Therefore in what follows we briefly review only the
safe and certain result of  \cite{LQG-D4}.

For the sake of simplicity we consider only six-derivative models,
as it was done in \cite{LQG-D4}. It proves useful to write the
six derivative action in a slightly different form,
\beq
S = - \frac{2}{\ka^2}
\int d^4x\sqrt{-g} R\,\,
-  \int d^4x\sqrt{-g}\,\Big\{
\frac{\al}{2}\,C_{\mu\nu\al\be}
\Pi_2C^{\mu\nu\al\be}
+ \al\om\,R\Pi_0 R \Big\},
\eeq
where $\Pi_{0,2} = \Pi_{0,2}\big(\Box\big)=1+\,...$ are
polynomials of the first order. In the momentum representation
one can write
\beq
\Pi_2 (p^2)\,=\,1 + \frac{p^2}{2A_2}
\,,\qquad
\Pi_0(p^2)\,=\,1 + \frac{p^2}{2A_0}\,,
\eeq
where $A_0$ and $A_2$ are constants of the $mass^2$ - dimension.

The part of the action which is  quadratic in the perturbations,
$\,\ka h_{\mu\nu}=g_{\mu\nu}-\eta_{\mu\nu}$, has the form
\beq
S^{(2)}_{red}
&=&
-\,\int d^4x
\Big\{
\frac12\,h^{\mu\nu}\Big[\frac{\al\ka^2}{2}
\Pi_2\big(\pa^2\big)\pa^2-1\Big]\,\pa^2
\,P^{(2)}_{\mu\nu,\,\rho\si}\,h^{\rho\si}
\\
\nonumber
&+& h^{\mu\nu}\Big[\al\om\ka^2\Pi_0\big(\pa^2\big)\pa^2-1\Big]
\,\pa^2\,P^{(0-s)}_{\mu\nu,\,\rho\si}\,h^{\rho\si}
\Big\}\,,
\eeq
where
\beq
P^{(0-s)}_{\mu\nu,\,\rho\si}
=\frac13\,\theta_{\mu\nu}\,\theta_{\rho\si}
\,,\quad
P^{(2)}_{\mu\nu,\,\rho\si}
= \frac12\,\big(\theta_{\mu\rho}\,\theta_{\nu\si}
+ \theta_{\nu\rho}\,\theta_{\mu\si}\big)
- P^{(0-s)}_{\mu\nu,\,\rho\si}\,,
\eeq
are projectors of the spin-0 (scalar) and spin-2 (tensor) modes, and
\beq
\theta_{\mu\nu} = \eta_{\mu\nu} - \frac{\pa_\mu\pa_\nu}{\pa^2}.
\eeq

After the Wick rotation  the equations for the poles are
\beq
\al\Pi_2 (p^2)p^2\,=\,2M_P^2
\,,\quad
\al\om\,\Pi_0(p^2)p^2\,=\,M_P^2\,.
\eeq

Now, the solution for the tensor part (scalar sector can be
elaborated in a similar way) is
\beq
p^2 = m_2^2 = - A_2 \pm \sqrt{A_2^2 + \frac{4A_2M_P^2}{\al}}\,.
\eeq

One can distinguish two possible cases in this solution.
\vskip 1mm

{\it i)} \
 Two real positive solutions $0<m_{2+}^2<m_{2-}^2$;
\vskip 1mm

{\it ii)} \
 Two pairs of complex conjugate solutions for the mass.
\vskip 1mm

In the theory of the field $h_{\al\be}$, the condition of unitarity of
the $S$-matrix can be formulated in a usual way,
\beq
S^\dagger S = 1 \,,
\quad
\mbox{or}
\quad
S = 1 + i T
\quad
\mbox{and}
\quad
- i ( T - T^\dagger) = T^\dagger T\,.
\eeq
By defining the scattering amplitude as
\beq
\langle f | T | i \rangle
\,=\,
(2 \pi)^D \,\,\delta^{D}(p_i - p_f) \, T_{f i}
\eeq
we arrive at
\beq
- i \left( T_{f i} - T^*_{i f} \right)
=
\sum_k  T^*_{k f} T_{k i}\,.
\eeq
If we assume that for the forward scattering amplitude $i = f$,
the previous equation simplifies to
\beq
2 \,  {\rm Im}\, T_{ii}
=
\sum_k T^*_{i k} \, T_{ik}  > 0 \,.
\label{4}
\eeq
The detailed analysis of tree-level, one-loop and multi-loop diagrams
shows that the relation (\ref{4}) is satisfied because massive poles
always show up in complex conjugate pairs. The analysis performed
in the reference \cite{LQG-D4} is mainly at the tree-level, but the
complete proof of unitarity can be done on the basis of  the $O(N)$
scalar model within the Lee-Wick approach, that was considered in
\cite{LW}, and especially in \cite{Cutk} and \cite{GCW}. The
proofs of  \cite{GCW} directly apply to the higher derivative gravity
superrenormalizable QG with complex massive poles. Finally we
can conclude that this QG theory is unitary, but there may be a
violation of causality at the microscopic time scales, defined by
the magnitude of masses.

\section{Ghost-induced instabilities in cosmology}
\label{s6}

The unitarity of the $S$-matrix can not be regarded as the
unique condition of consistency of the QG theory. Even more
than that:  since gravity is essentially a non-polynomial theory,
unitarity can not be seen even as the most relevant consistency
condition. The main requirement should be the stability of
physically relevant solutions of classical general relativity in
the presence of higher derivatives and massive ghosts.

The study of stability of the general gravitational solutions in the
presence of higher derivatives does not look a realistic problem
to solve. There are a few publications \cite{Whitt,Myung} (see
also \cite{Aux}) with  conflicting results concerning the stability
of Schwarzschild solution in fourth order gravity. The study of
this subject is very complicated and can not be described in this
short review. Hence we will concentrate on the stability on the
cosmological background which is much better explored.

The problem has been explored in several old and newer
publications, for different cosmological backgrounds. In the
case of gravitational waves on de Sitter space  and the
typical energy of the wave much below $\,M_p$ the situation
was described in \cite{star81,hhr,anju} and in a more detailed
and elaborated form, with the special attention to the role of
higher derivatives,  in \cite{GW-Stab}. Recently, the case of
more general cosmological backgrounds has been reported
in Ref.~\cite{HD-Stab} (see also a short review in
\cite{GW-HD-MPLA}). Let us start by explaining these
results.

\subsection{Perturbations: low values of $k$}

\begin{figure}[H]
\centering
\includegraphics[scale=0.7]{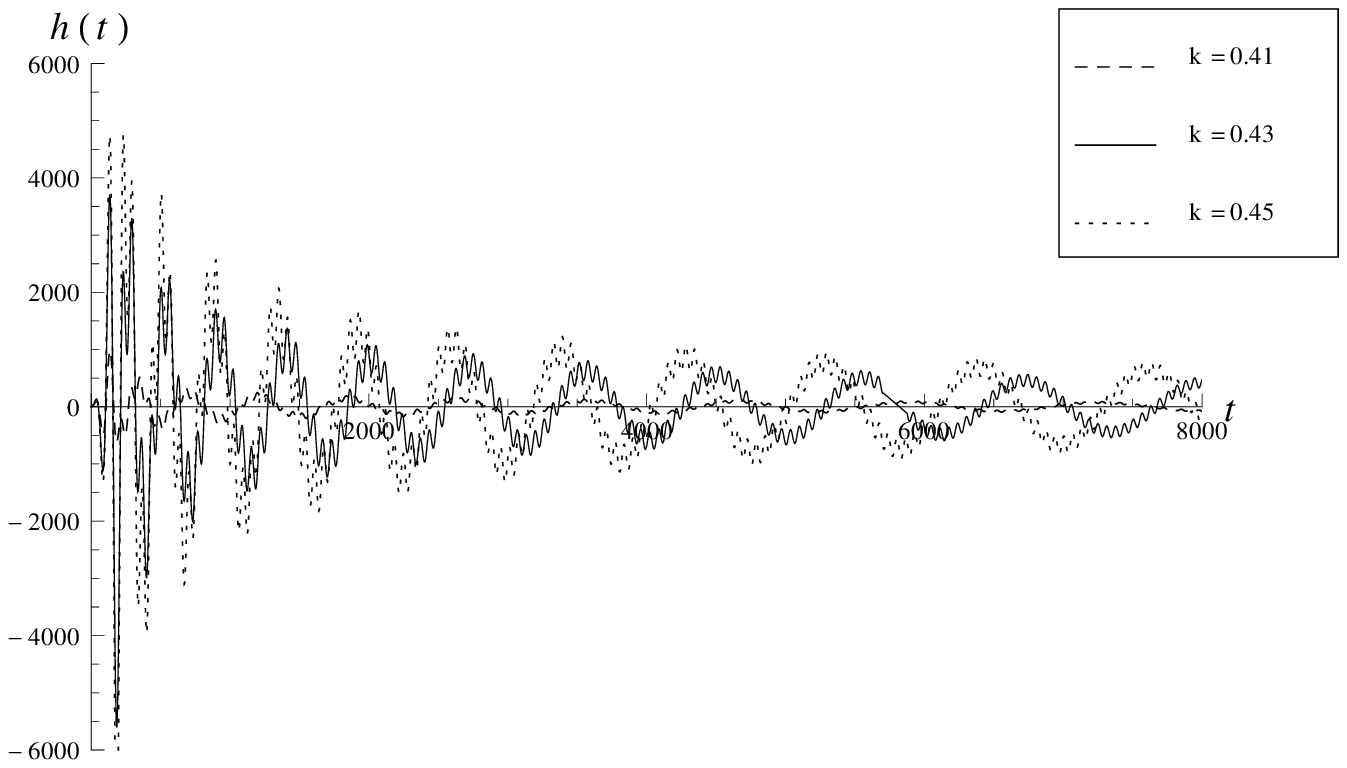}
\includegraphics[scale=0.7]{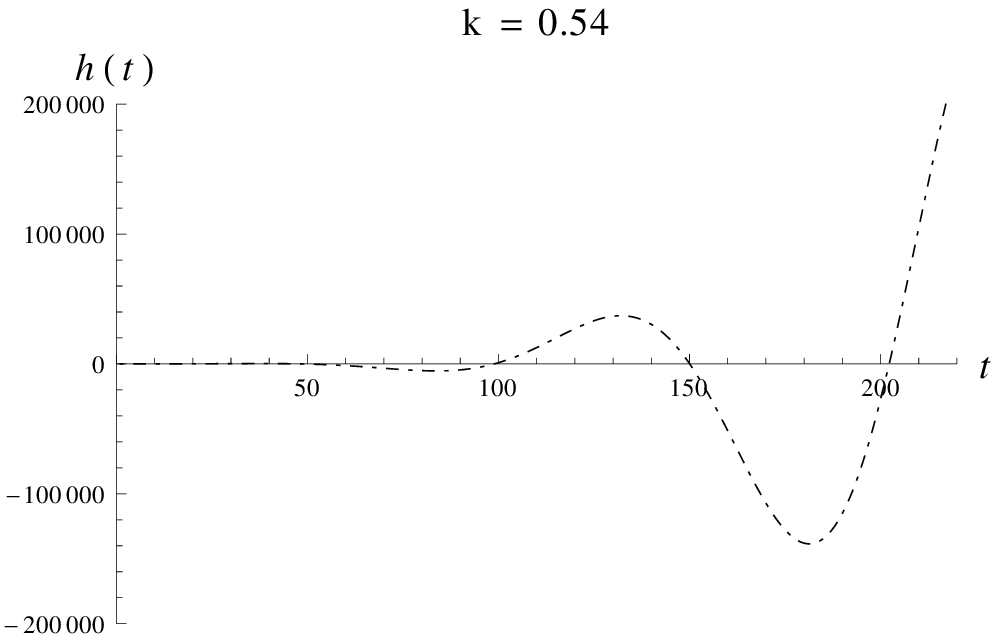}
\caption{The solution with growing modes appear only
starting from $k=0.54\,M_P$}
\label{figure1}
\end{figure}

The main conclusion of \cite{HD-Stab} was that the absence of
growing modes in the fourth derivative theory holds if only if the
initial seeds of the gravitational perturbations have frequencies
below the threshold which is of the order of Planck mass. The
situation is illustrated in FIG.~\ref{figure1} for the specific case
of radiation-dominated Universe.

One can observe in  FIG.~\ref{figure1} that there are no growing
modes, until the frequency $k$ achieves the value $\approx 0.54$ in
the Planck units.  Starting from this value, we observe instability
due to the effect of massive ghost. Our interpretation of this result
is that the ghost is present in the spectrum of the theory, but if
there is no real ghost in the universe, there is no instability. The
massive ghost can not be created from the vacuum if the density
of gravitons does not approach the Planck density which is required
to create a ghost from the vacuum state.

Of course, the concentration of gravitons of the Planck order of
magnitude is not forbidden by all known physical laws. Hence
we can expect that some new laws should be discovered to
resolve the problem of consistent QG. And we can see that these
new laws must forbid the Planck order density of gravitons to
resolve the issue, at least for the case of a cosmological
background.

Let us note that the semiclassical (anomaly-induced) corrections
were also included into consideration \cite{HD-Stab}. As far as these
 corrections ate at least ${\cal O}(R^3_{....})$, it is natural that
 the qualitative result for the Planck order threshold for stability
 does not change.  The reason is that until the energy of the
 gravitational  perturbations does not approach the Planck order of
 magnitude, these corrections can not compete with the classical
 ${\cal O}(R^2_{....})$-terms and, e.g., their running.

\begin{figure}[H]
\centering
\includegraphics[scale=0.8]{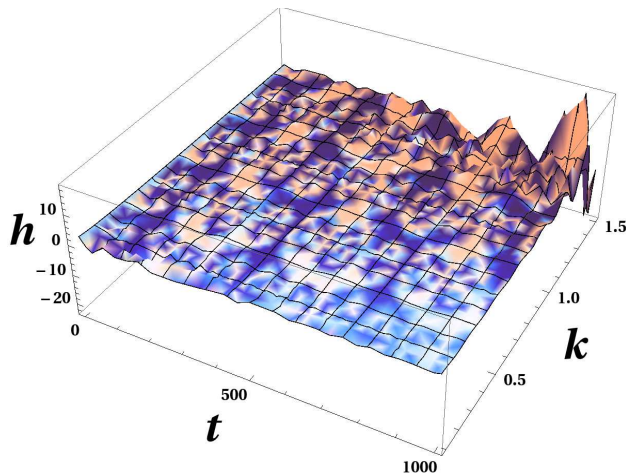}
\hspace{1cm}
\caption{In this plot we are using the units $M_P=1$ and the
values $a_1 = -1$ and $H=M_P$. The tensor perturbation
mode $h$ is shown as function of time $t$ and of the initial
frequency $k$. Until the values of $k$ are small there are
no strongly growing solutions. But when $k$ gets closer to the
Planck scale, the perturbations start to explode because of
the high derivatives terms.}
\label{figure3}
\end{figure}

In order to illustrate better the existence of the Planck threshold,
we included the $3D$ plot in FIG.~\ref{figure3}.

In this figure one can observe perturbation $h$ as function
of time and of the initial frequency $k$.
In the 3D plot one can observe a ``normal'' oscillation for small values
of $k$, and then the solution explodes abruptly for $k$ close to the
Planck mass scale. Apparently, for the values $k>M_P$ there are
run-away solutions.

\subsection{Perturbations: high values of $k$}

Thus we have a generally optimistic situation for the sub-Planckian
frequencies. Indeed, this is not a really nice situation, from the general
perspective. The remaining question is what can  we do with ghosts
in the case of Planck order or greater frequencies? To answer this
question let us follow \cite{Salles:2017xsr} and take a look at the
simplest possible equation for the fourth-derivative gravity without
quantum or semiclassical corrections,
\beq
\frac{1}{3}  h^{\lp\textsc{\tiny IV}\rp}
+ 2H h^{(\textsc{\tiny III})}
+ \Big( H^2 + \frac{\MP^2}{32\pi a_1}\Big) \ddot{h}
+  \frac{1}{6} \frac{\nabla^4 h}{a^{4}}
- \frac{2}{3}\frac{\nabla^2 \ddot{h}}{a^2}
-  \frac{2H}{3} \frac{\nabla^{2} \dot{h}}{a^{2}}
\nonumber
\\
- \Big( H \dot{H} + \ddot{H} + 6 H^3
 - \frac{3\MP^2 H}{32\pi a_1}\Big) \dot{h}
 - \Big[ \frac{\MP^2}{32\pi a_1}
- \frac43  \lp \dot{H} + 2H^2\rp
\Big] \frac{\nabla^2 h}{a^{2}}
\nonumber
\\
- \Big[
24 \dot{H} H^{2} + 12 \dot{H}^{2} + 16 H \ddot{H}
+ \frac83 H^{\lp\textsc{\tiny III}\rp}
-\frac{\MP^{2}}{16\pi a_1}\lp 2 \dot{H} + 3 H^{2}\rp\Big] h
 = 0.
\nonumber
\eeq
It is easy to note that the space derivatives $\na$ and hence the
wave vector ${\vec k}$ enter this equation only in the combination
\beq
{\vec q} \,=\, \frac{{\vec k}}{a(t)}.
\eeq
When the universe expands, the frequency becomes smaller.  This
qualitative conclusion is supported by numerical analysis described
in Ref.~\cite{Salles:2017xsr}, including the model with semiclassical
corrections taken into account.

\begin{figure}[H]
\centering
\includegraphics[scale=0.56]{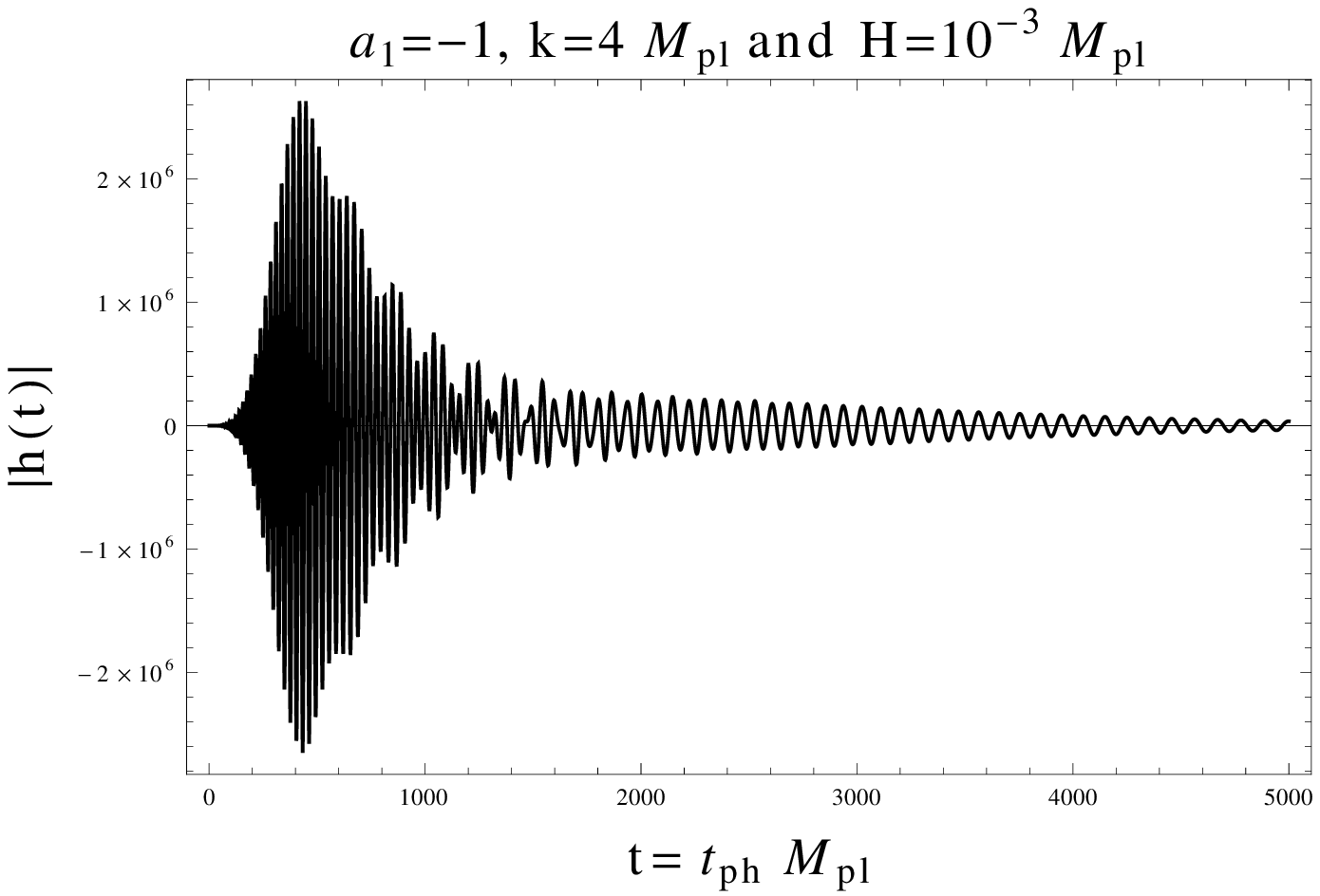}
\includegraphics[scale=0.56]{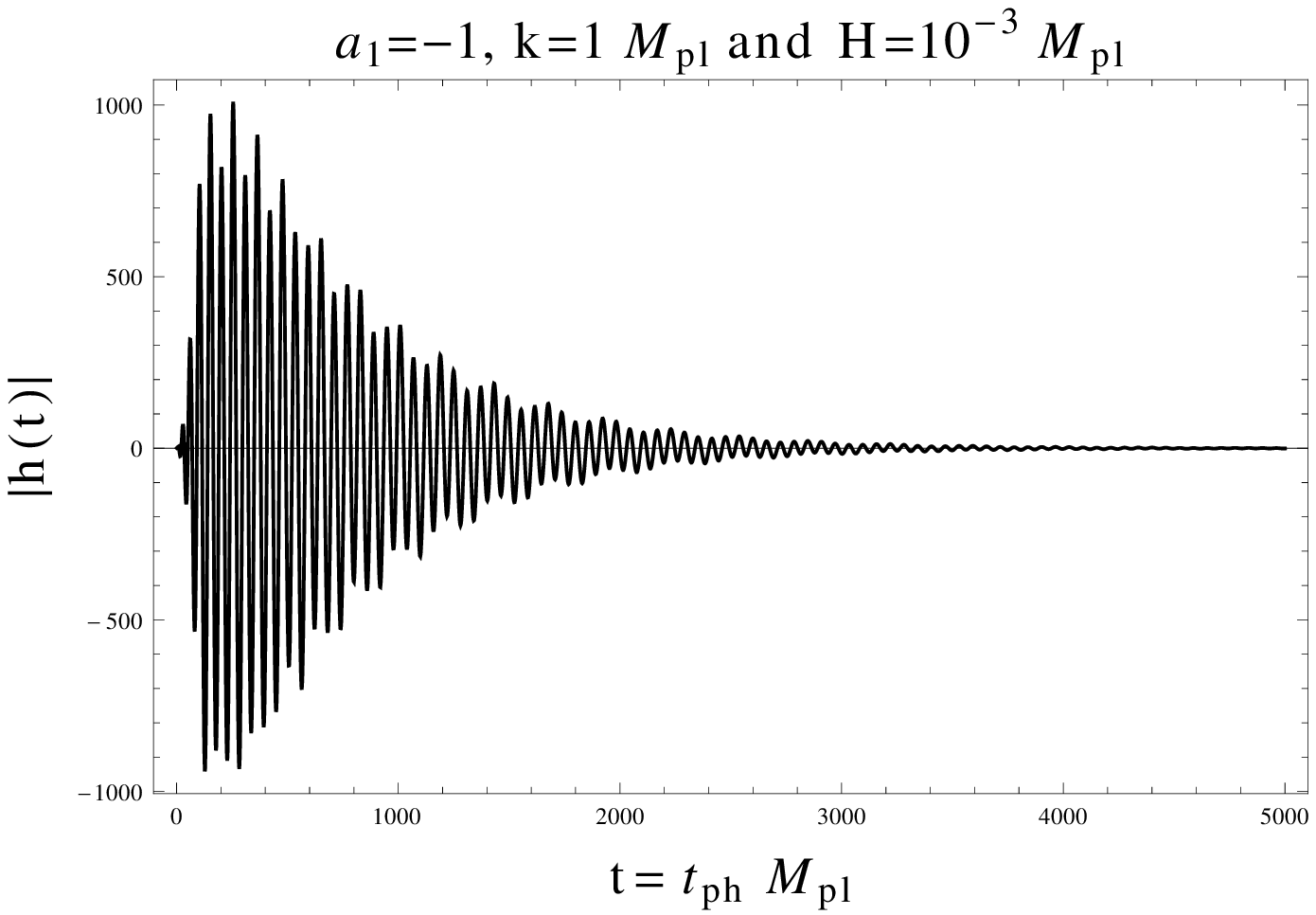}
\caption{In the case of radiation $\,a \sim \sqrt{t}\,$ background
and transplanckian frequencies there is an explosive growth of
perturbations, that stabilize soon after it starts. In these two plots
we are using $\MP$ (normalized), $a_1 = -1$, $H=10^{-3} \MP$
and $k=4,1$.}
\label{figure2}
\end{figure}

In FIG.~\ref{figure2} one can see that the growth of the
gravitational waves with transplanckian frequencies really stops
at some point. At least in the cosmological setting this may be a
solution of the general problem.

In this case we have $\,k \geq M_P\,$ in the gravitational theory
with high derivatives terms. Unlike the previous case of relatively small
frequencies one can observe the effects of ghosts, since the run-away
solutions almost instantly appear. However, after a while these solutions
get damped, because the effective frequency tends to decrease due to
the fast expansion of the universe.

\section{Conclusions}
\label{s8}

In conclusions, let us make a few statements about the situation
with ghosts which was described above.
\vskip 1mm

{\it i)} \
We know that there is no way to have semiclassical or quantum
gravity without higher derivatives. The effective approaches imply
treating higher derivatives as small perturbations over the basic
theory which is GR with the cosmological constant. However, this
treatment has several weak points. First of all, is it completely
{\it ad hoc} and does not follow from the QFT logic, quite
different from the situation in QED, where higher derivatives
emerge only in the loop corrections, the corresponding terms do
not run and treating them as small corrections does not lead to
inconsistencies at the energy scales  where the theory is supposed
to work. The situation in QG is completely different, because the
 last is supposed to apply up to the Planck energies.
\vskip 1mm

{\it ii)} \
Higher derivatives mean ghosts and instabilities. But in the
closed system the problem can be solved because there is no
energy to provide a global and total explosion of ghost or even
tachyonic ghost modes (Lee-Wick approach). This way of
thinking provides the theory which is formally superrenormalizable
and unitary at the same time. However, this \ {\it does not} \ solve
the problem of stability, which remains open. The main reason
is that the real gravitational systems are not closed, and the
metric perturbations propagate on the non-trivial backgrounds
of classical gravitational solutions. Therefore one needs an
essential completion, or supplement to the proof of unitarity. This
means we need a new insight about how the stability problem can
be solved.
\vskip 1mm

{\it iii)} \
The analysis of linear stability on the cosmological background
shows that the perturbations with the initial seeds with the
frequencies below the Planck-order threshold do not grow.
The natural interpretation of this fact is that without the
Planck-order density of gravitons one can not create ghost
from the vacuum.

Maybe there is some unknown principle of Physics which
forbids Planck-scale concentration of gravitons? Some
discussion of the physical consequences of such a principle
has been recently discussed in the literature \cite{DG}.

The restriction on the initial frequencies can be violated
for the Planck-scale background, which ``opens'' the phase
space of quantum  states and enables the production of
instabilities. But after that the expansion of the universe
reduce the frequencies and the instabilities get stabilized.
This specific behaviour of perturbations in the theories
with higher derivatives creates a hope to observe the traces
of these theories in observations of transplanckian effects,
as it was discussed in \cite{BranMart,StarTransPlanck}.

\begin{acknowledgments}
This review is based on the original works which were done in
collaboration with Leonardo Modesto and Patrick Peter.
We gratefully acknowledge these collaborations.
I.Sh. is grateful to CAPES, CNPq and FAPEMIG for partial
support of his work.
\end{acknowledgments}






\begin{thebibliography}{99}

\bibitem{HD-Stab} F. de O. Salles and I.L. Shapiro,
Phys. Rev. D {\bf 89}, 084054 (2014);
%
{\bf 90}, 129903 (2014) [erratum],
\ arXiv:1401.4583.

 \bibitem{Salles:2017xsr}
  P.~Peter, F.D.O.~Salles and I.L.~Shapiro,
  Phys. Rev. {\bf D97} (2018) 064044,
arXiv:1801.00063.

\bibitem{GWprT}
G. Cusin, F.O. Salles and I.L. Shapiro,
Phys. Rev. {\bf D93} (2016) 044039, arXive:1503.08059.

 \bibitem{birdav} N.D. Birell and P.C.W. Davies,
{\it Quantum Fields in Curved Space}
(Cambridge University Press, Cambridge, 1982).

\bibitem{book}
I.L. Buchbinder, S.D. Odintsov and I.L. Shapiro,
{\it Effective Action in Quantum Gravity}.
IOP Publishing, Bristol (1992).

\bibitem{PoImpo} I.L. Shapiro,
Class. Quant. Grav. {\bf  25}, 103001 (2008). arXiv:0801.0216

\bibitem{hove}
G. t'Hooft and M. Veltman,
Ann. Inst. H. Poincare {\bf A20} (1974) 69.

\bibitem{dene} S. Deser and P. van Nieuwenhuisen,
Phys. Rev. {\bf 10D}, (1974) 401-410.

\bibitem{gorsag}
M. H. Goroff  and A. Sagnotti,
Nucl. Phys. {\bf B266} (1986) 709. 

\bibitem{Don94} J. Donoghue,
Phys. Rev. Lett. {\bf 72}, 2996 (1994);
Phys. Rev. {\bf D50} (1994)  3874. 

\bibitem{Burgess} C.P. Burgess,
Living Rev. Rel. {\bf 7} (2004) 5, 
gr-qc/0311082.

\bibitem{Stelle}
K.S. Stelle, Phys. Rev. {\bf D16} (1977) 953.

\bibitem{Ostrog} M.V.~Ostrogradsky,
{\it M\'emoires sur les \'equations différentielles, relatives
au problème des isopérimètres},
Mem. Acad. St. Petersbourg,
{\bf 6} (1850) 385. 

\bibitem{Woodard-r} R.P. Woodard,
Lect. Notes Phys. {\bf 720} (2007) 403, astro-ph/0601672

\bibitem{Veltman-63}  M. J. G. Veltman,
 Physica {\bf 29} (1963) 186. 

\bibitem{Tomb-77} E.~Tomboulis,
Phys. Lett. B {\bf 70} (1977) 361;
 {\bf 97} (1980) 77;
Phys. Rev. Lett. {\bf 52} (1984) 1173.

\bibitem{salstr} A.~Salam and J.~Strathdee,
Phys. Rev. D {\bf 18} (1978) 4480.

\bibitem{Hawking}
S.W.~Hawking and Th.~Hertog,
Phys. Rev.  {\bf D65} (2002) 103515.

\bibitem{highderi} M. Asorey, J.L. L\'opez, I.L. Shapiro,
Int. Journ. Mod. Phys. {\bf A 12} (1997) 5711, hep-th/9610006.

\bibitem{Newton-high} L. Modesto, T. de Paula Netto, I.L. Shapiro,
JHEP {\bf 1504} (2015) 098, arXiv:1412.0740;
\\
B.L. Giacchini,
Phys. Lett. {\bf B766} (2017)  306.

\bibitem{ABS}
  A.~Accioly, B.L.~Giacchini and I.L.~Shapiro,
  Phys. Rev.  {\bf D96} (2017) 104004,
arXiv:1610.05260;
Eur. Phys. J. {\bf C77} (2017) 540
arXiv:1604.07348.

\bibitem{SRQG-beta} L.~Modesto, L.~Rachwal and I.L.~Shapiro,
Eur. Phys. J. {\bf C78} (2018) 
555,
arXiv:1704.03988.

\bibitem{Simon-90} J.Z. Simon,
Phys. Rev. {\bf D41} (1990) 3720;
Phys. Rev. {\bf D43} (1991) 3308; 
\\
Phys. Rev. {\bf D45} (1992) 1953. 

\bibitem{parsim}
L. Parker, J.Z. Simon,
Phys. Rev. {\bf D47} (1993) 1339, 
gr-qc/9211002.

\bibitem{BSSh-NPB}
I.L. Buchbinder, I.L. Shapiro and  A.G. Sibiryakov,
Nucl. Phys. {\bf B445} (1995) 109. 

\bibitem{FrTs-85}
E.S. Fradkin and A.A. Tseytlin , Phys. Lett. {\bf B158} (1985)  316;
Nucl. Phys. {\bf B261} (1985) 1.

\bibitem{CFMP} C. Callan, D. Friedan, E. Martinec and M. Perry,
Nucl. Phys. {\bf B272} (1985) 593.

\bibitem{zwei}
B.~Zwiebach, Phys. Lett. B {\bf 156} (1985) 315;
\\
S.~Deser and A.N.~Redlich, \textit{ibid.} {\bf 176} (1986) 350;
\\
A.A.~Tseytlin, \textit{ibid.} {\bf 176} (1986) 92.

\bibitem{maroto}
A.L. Maroto and I.L. Shapiro, Phys. Lett. B {\bf 414} (1997) 34.

\bibitem{star}
A.A.~Starobinsky, Phys. Lett. {\bf B91} (1980) 99.

\bibitem{Tseytlin-95} A.A.~Tseytlin,
Phys. Lett. B {\bf 363} (1995) 223, \ 
hep-th/9509050.

\bibitem{CM} G. Calcagni, L. Modesto, and G. Nardelli,
{\it  Nonperturbative spectrum of nonlocal gravity,}
arXiv:1803.07848;
JHEP 1805 (2018) 087
arXiv:1803.00561.

\bibitem{Tomboulis-97} E.T. Tomboulis,
{\it Superrenormalizable gauge and gravitational theories},
hep-th/9702146;
Phys. Rev. {\bf D92} (2015) 125037,
arXiv:1507.00981; \
Mod. Phys. Lett. {\bf A30} (2015) 1540005.

\bibitem{Modesto2011}
  L.~Modesto,
  Phys. Rev.  {\bf D86} (2012) 044005, arXiv:1107.2403.

\bibitem{CountGhost} I. L. Shapiro,
Phys. Lett. {\bf B744} (2015) 67,
arXiv: 1502.00106.

\bibitem{UtDW}
R. Utiyama and B.S. DeWitt, J. Math. Phys. 3 (1962) 608.

\bibitem{Antomb} 	
I.~Antoniadis and E.T.~Tomboulis,
Phys. Rev. D {\bf 33} (1986) 2756.

\bibitem{Johnston} 	
D.~A.~Johnston,
Nucl. Phys. B {\bf 297} (1988) 721.

\bibitem{RobertoP}  A. Codello, and R. Percacci,
Phys. Rev. Lett. {\bf 97} (2006) 221301,
hep-th/0607128.

\bibitem{LQG-D4} L. Modesto and I.L. Shapiro,
Phys. Lett.  {\bf B755} (2016)  279, arXiv:1512.07600.

\bibitem{Modesto2016} L. Modesto,
Nucl. Phys. {\bf B909} (2016) 584, 
arXiv:1602.02421.

\bibitem{ARS}  M.~Asorey, L.~Rachwal and I.L.~Shapiro,
Galaxies {\bf 6} (2018) 23,
arXiv:1802.01036.

\bibitem{MOD} M.~Christodoulou and L.~Modesto,
{\it Reflection positivity in nonlocal gravity,}
arXiv:1803.08843.

\bibitem{LW}
  T.~D.~Lee and G.~C.~Wick,
  Phys. Rev.  {\bf D2} (1970) 1033;
  %
  T.~D.~Lee and G.~C.~Wick,
  Nucl. Phys. {\bf B9}  (1969) 209.

\bibitem{Cutk}
  R.E.~Cutkosky, P.V.~Landshoff, D.I.~Olive and J.C.~Polkinghorne,
  Nucl. Phys. {\bf B12} (1969) 281.

\bibitem{GCW} B. Grinstein, D. O'Connell and M. B. Wise,
Phys. Rev. {\bf D79} (2009) 105019,
hep-th/0805.2156.

\bibitem{Whitt}
B.~Whitt,
{\it Phys. Rev.} {\bf D32}  (1985) 379. 

\bibitem{Myung} Yu.S.~Myung,
{\it Phys. Rev.} {\bf D88}  (2013) 024039, arXiv:1306.3725.

\bibitem{Aux} S. Mauro, R. Balbinot, A. Fabbri
and .L. Shapiro,
Europ. Phys. Journ. Plus {\bf 130} (2015) 135,
arXiv:1504.06756.

\bibitem{star81}
A.A. Starobinsky,
Zh. Eksp. Teor. Fiz. {\bf 34} (1981) 460. 

\bibitem{hhr} S.W. Hawking, T. Hertog and H.S. Real,
Phys.Rev. {\bf D63} (2001) 083504.

\bibitem{anju} J.C.Fabris, A.M.Pelinson and I.L.Shapiro,
Nucl. Phys. {\bf B597} (2001) 539.

\bibitem{GW-Stab} J.C. Fabris, A.M. Pelinson, F. de O. Salles, and
I.L. Shapiro,
JCAP {\bf 02} (2012) 019; \ arXiv: 1112.5202.

\bibitem{GW-HD-MPLA}
I.L. Shapiro, A.M. Pelinson, and F. de O. Salles,
Mod. Phys. Lett. {\bf A29} (2014) 1430034,
arXiv:1410.2581.

\bibitem{DG}
G. Dvali, S. Folkerts, and C. Germani,
Phys. Rev. D {\bf 84} (2011) 024039;  
%
G. Dvali, and C. Gomez,
Fortschr. Phys. {\bf 63} (2013) 742, arXiv:1112.3359.

\bibitem{BranMart} J. Martin, and R.H. Brandenberger,
Phys. Rev. {\bf D63} (2001) 123501,
hep-th/0005209.

\bibitem{StarTransPlanck} A.A. Starobinsky,
Pisma Zh. Eksp. Teor. Fiz. {\bf 73} (2001) 415 (in Russian),
English translation: JETP Lett. 73 (2001) 371,
astro-ph/0104043.

\end{thebibliography}
\end{document}